\documentclass[aps,prl,twocolumn,superscriptaddress,raggedbottom,showpacs,floatfix]{revtex4-1}
\usepackage{amsmath,amsfonts,amssymb,amsthm,graphics,graphicx,latexsym,color}
\usepackage[next]{inputenc}
\usepackage[dvips]{epsfig}
\usepackage[colorlinks=true,citecolor=blue,linkcolor=blue]{hyperref}
\usepackage{bbm,bm, bbold}
\usepackage{booktabs}
\usepackage{multirow}
\usepackage{hhline}
\usepackage{comment}
\usepackage{float}

\usepackage{epstopdf}

\usepackage{natbib}

\begin{document} 
\renewcommand{\vec}{\mathbf}
\renewcommand{\Re}{\mathop{\mathrm{Re}}\nolimits}
\renewcommand{\Im}{\mathop{\mathrm{Im}}\nolimits}

\title{Exciton-polarons in doped semiconductors in a strong magnetic field}
\author{Dmitry K. Efimkin}
\affiliation{Center for Complex Quantum Systems, University of Texas at Austin, Austin Texas 78712, USA}

\author{Allan H. MacDonald}
\affiliation{Center for Complex Quantum Systems, University of Texas at Austin, Austin Texas 78712, USA}

\begin{abstract}
In previous work we have argued that the optical properties of moderately doped two-dimensional semiconductors
can be described in terms of excitons dressed by their interactions with a degenerate Fermi sea of additional charge carriers. These interactions split the bare exciton into attractive and repulsive exciton-polaron branches. 
The collective excitations of the coupled system are many-body generalizations of the bound trion and 
unbound states of a single electron interacting with an exciton.   
In this article we consider exciton-polarons in the presence of an 
external magnetic field that quantizes the kinetic energy of the electrons in the Fermi sea.
Our theoretical approach is based on a transformation to new basis that respects the underlaying symmetry of magnetic
translations.   We find that the attractive exciton-polaron branch is only weakly influenced by the magnetic field,
whereas the repulsive branch exhibits magnetic oscillations and splits into discrete peaks
that reflect combined exciton-cyclotron resonance.  

\end{abstract}
\maketitle

\noindent

\section{I. Introduction}
The monolayer transition metal dicholagenides (TMDC), $\hbox{MoS}_2$, $\hbox{MoSe}_2$, $\hbox{WS}_2$ and $\hbox{WSe}_2$, are widely studied 
beyond-graphene~\cite{Graphene1, Graphene2, GrapheneMacDonald} two-dimensional semiconductors~\cite{TMDC1,TMDC2,TMDC4,TMDC5,TMDC6}. 
Strong Coulomb and spin-orbit interactions, tunable optical properties, and valley-selection via circularly polarized
light, combine to make them promising materials for optoelectronics and valleytronics~\cite{TMDCReview1,TMDCReview2}. 
TMDC monolayers and heterostructures have therefore emerged as 
an interesting platform for the exploration of 
exciton physics (See Ref.~\cite{TMDCExcitonReview} and references therein).

Recent experiments have demonstrated that the absorption spectra of TMDC monolayers is dominated by an exciton features with a large binding energies $\epsilon_\mathrm{X}\approx 0.5\;\hbox{eV}$~\cite{TMDCEx1,TMDCEx2}. 
When charge carriers are present the exciton absorption feature splits into two separate peaks~\cite{TrionExperiment1,TrionExperiment2,TrionExperiment3,TrionExperiment4, TrionExperiment5, TrionExperiment6,TrionExperiment7,TMDCTrionEn1,TMDCTrionEn2,TMDCValley}, that are usually 
attributed to excitons ($\mathrm{X}$) and to  trions ($\mathrm{T}$), charged weakly-coupled three-particle complexes formed by binding two electrons to one hole or two holes to one electron. 
The difference between resonant frequencies for the peaks is interpreted as the binding energy $\epsilon_\mathrm{T}$ of a trion. This phenomenon has been observed previously in conventional quantum well 
systems based on $\hbox{GaAs}$ and $\hbox{CdTe}$~\cite{QW1, QW2, QW3, QW4, QW5}, 
although the binding energy of trions there $\epsilon_\mathrm{T}\approx 2~\hbox{meV}$  is more than an order of magnitude smaller than in TMDC $\epsilon_\mathrm{T}\approx 30~\hbox{meV}$~\cite{TMDCTrionEn1,TMDCTrionEn2}. In our previous work~\cite{EfimkinMacDonald} we have argued that the 
three-particle picture of the additional peak is appropriate only at low doping 
$\epsilon_\mathrm{F}\ll\epsilon_\mathrm{T}$, where $\epsilon_\mathrm{F}$ is the Fermi energy of excess charge carriers. 
The three-particle picture cannot explain the competition for spectral weight 
between two peaks that is observed in experiments at higher carrier densities (See also related earlier work~
\cite{Suris1, Suris2, Wouters1, Wouters2, Combescot1, Combescot2, Combescot3}). In the wide doping range
where  
$\epsilon_\mathrm{F}\lesssim\epsilon_\mathrm{T}$, that in TMDC monolayers corresponds to concentrations $n\lesssim 5 \;10^{12}\; \hbox{cm}^{-2}$, the appropriate picture is instead one of renormalized excitons interacting with the degenerate Fermi sea of excess charge carriers~\cite{EfimkinMacDonald,TMDCDemlerExp}. As in the 
Fermi-polaron problem~\cite{FermiPolaron1,FermiPolaron2, FermiPolaron3,FermiPolaron4,FermiPolaron5, FermiPolaronReview1,FermiPolaronReview2,FermiPolaronReview3}, the interactions dress excitons into exciton-polarons. Exciton-polarons have attractive and repulsive spectral branches that evolve from, and generalize to degenerate carrier densities, the separate absorption processes associated with
bound and unbound trion states.

Here we consider exciton-polarons in the presence of magnetic field of the moderate strength $\hbar \omega_\mathrm{B}\sim\epsilon_\mathrm{T}$, where $\hbar \omega_\mathrm{B}$ is the spacing between Landau levels for excess charge carriers.  In this regime the magnetic field is too weak to alter the structure of 
excitons~\cite{Magnetoexciton2,Magnetoexciton3}. On the other hand the excess charge carriers are in the regime of the integer and fractional quantum Hall effect, and the polaron dressing of excitons can be seriously influenced. In TMDC materials with the binding energy $\epsilon_\mathrm{T}\approx 30~\hbox{meV}$~\cite{TMDCTrionEn1} the polaronic dressing of excitons is strongly modified at magnetic field of order $B\sim 85 \hbox{T}$, while magnetic oscillations of the repulsive exciton-polaron branch we present below could be observable at smaller fields. Optical studies of TMDC at magnetic field $B\approx 10\;\hbox{T}$, not enough strong to influence the polaron dressing, have been presented~\cite{TMDCMF1, TMDCMF2, TMDCMF3, TMDCMF4, TMDCMF5}. Studies of undoped TMDC samples at much stronger magnetic fields $B\approx 65\;\hbox{T}$ have been reported 
very~\cite{TMDCStrongMF} recently, and we anticipate that experiments with doped samples at such fields will be 
undertaken soon. 

\begin{figure}[t]
	\label{Fig1}
	\includegraphics[width=0.83\columnwidth]{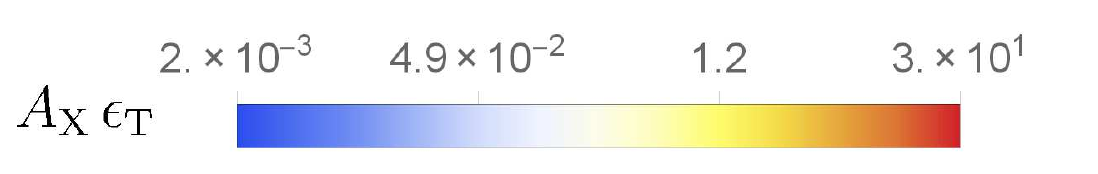}
	\vspace{-2 pt}
	\includegraphics[width=0.85\columnwidth]{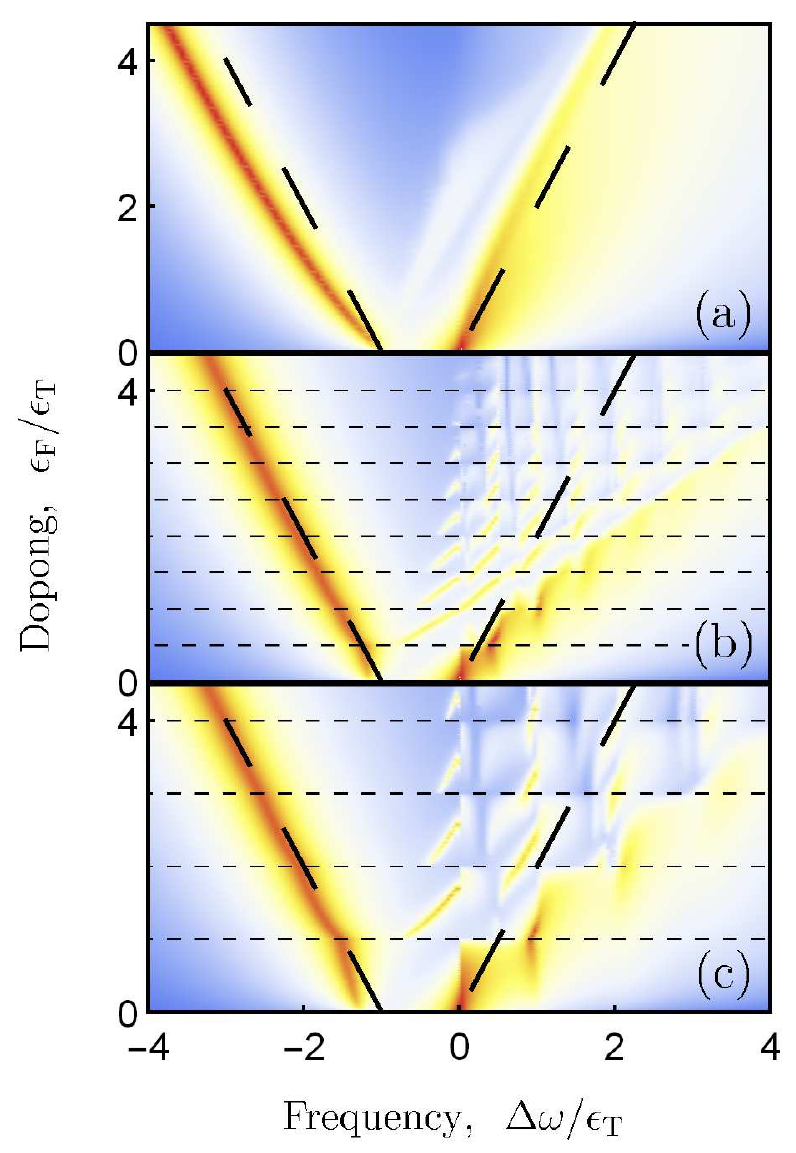}
	\vspace{-2 pt}
	\caption{The spectral function of excitons at zero momentum $A_\mathrm{X}(\omega,0)$ 
	in the absence of a magnetic field $\hbox{(a)}$. The excitonic feature splits into attractive and repulsive exciton-polaron branches,
		with dispersions that can be approximated by $\epsilon_\mathrm{T}^*=-3\epsilon_\mathrm{T}/2-\epsilon_\mathrm{F}/2$\, and $\epsilon_\mathrm{X}^*=\epsilon_\mathrm{F}/2$.  The approximate dispersions are plotted as long-dashed lines. The spectral functions $A_\mathrm{X}(\omega,0)$ are illustrated
		 at magnetic fields of strength $\hbar\omega_\mathrm{B}/\epsilon_\mathrm{T}=0.5$~$\hbox{(b)}$ 
		 and $1$~$\hbox{(c)}$. The horizontal short-dashed lines denote integer filling factors $\nu_\mathrm{e}$.  The repulsive exciton-polaron branch splits into a number of peaks, which can be interpreted as
		  combined exciton-cyclotron resonances.} 
\end{figure} 

In our theory the optical conductivity is proportional to the spectral function of excitons with zero momentum $A_\mathrm{X}(\omega,0)$.  Our main results for $A_\mathrm{X}(\omega,0)$ are 
summarized in Fig. 1. The splitting of excitons into attractive and repulsive exciton-polaron branches in the absence of magnetic field is illustrated in Fig.1~$\hbox{(a)}$. 
The spectral weight  $A_\mathrm{X}(\omega,0)$ at  magnetic field strengths $\hbar \omega_\mathrm{B}/\epsilon_\mathrm{T}=0.5$ and $1$ are presented in Fig.1~$\hbox{(b)}$ and~$(\hbox{c})$. 
The attractive exciton-polaron branch is only weakly influenced by the magnetic field. The repulsive branch splits into a number of peaks separated by the Landau quantization energy $\hbar \omega_\mathrm{B}$, 
and experiences magnetic-oscillations as the Fermi level passes through different Landau levels.  The absorption peaks that are blueshifted from the primary repulsive exciton-polaron peak can be interpreted as combined exciton-cyclotron resonances and correspond to the simultaneous creation of an exciton and an electronic transition between Landau levels. With decreasing magnetic field the number of additional peaks increases 
and the absorption pattern smoothly evolves into the dispersive the zero-field repulsive exciton-polaron branch.

Our paper is organized as follows. In Sec. II we introduce the model we employ to describe the optical properties of TMDCs. In Sec. III some pertinent aspects of  electronic structure theory in a magnetic field are reviewed. In Sec. IV. some features of the two-body exciton-electron problem are discussed. Sec. V is devoted to the many-body exciton-polaron problem. In Sec. VI. we present and discuss results. Finally in Sec. VII we conclude, compare our results with experimental work at conventional semiconductors and comment on limitations of our theory. 

\section{II. The model}

The minimal model for the optical properties of single-layer TMDCs accounts for 
two valleys with conventional parabolic energy spectra~\footnote{The model we consider is the non-relatvistic limit of a four band model for massive Dirac particles with strong spin-valley coupling.  The role of Berry phases and the Dirac-like spectrum of TMDC semiconductors has been discussed in several recent papers~\cite{ExcitonDirac1,ExcitonDirac2,ExcitonDirac3,ExcitonDirac4,ExcitonDirac5, ExcitonDirac6}. They are important only for excited excitonic states which are excluded from considerations here}. Due to strong Coulomb interactions, absorption is dominated by excitation of direct excitons with an resonant energy $\omega_\mathrm{X}$ well below the single particle continuum threshold. Importantly, the two valleys can be probed independently by using circularly polarized light to achieve valley-selective absorption.

Our polaronic theory is based on the difference between trion and exciton binding energy scales,
{\it i.e.} on the property that $\epsilon_\mathrm{T}\ll \epsilon_\mathrm{X}$.  This separation 
allows us to account for the effect of excess charge carriers and magnetic fields 
by adding a polaronic dressing self-energy to the exciton propagator.
Provided that $\epsilon_\mathrm{F} \ll  \epsilon_\mathrm{X}$, the formation of photoexcited excitons is 
weakly disturbed, and they can still be treated as composite bosons with 
bare resonant energy $\omega_\mathrm{X}$, mass $m_\mathrm{X}$ and optical transitions matrix element $D$. 
We assume that the bare excitons interact with the excess charge carriers via a phenomenological contact 
pseudopotential that accounts approximately for 
their short-range monopole-dipole Coulomb interactions. 
Due to the presence of two valleys, excitons are disturbed by two Fermi seas. In our model, we neglect interactions between the exciton and carriers in the same valley which are weakened by the exchange physics related to
the possibility that electrons can be part of the exciton and in the Fermi sea at the same time~\footnote{The low-energy model describes excitonic transitions between low-energy bands of the TMDC four band electronic structure, which are referred as $\hbox{A}$ excitons. Excitonic transitions between high-energy bands, referred as $\hbox{B}$ excitons, can be dressed by Fermi seas from two valley since low-energy bands are filled by electrons and exchange interactions are absent. The corresponding valley splitting of the attractive exciton-polaron branch have been recently reported~\cite{TMDCMF5,TMDCTrionEn2,TMDCValley}.}. Interactions with the Fermi sea from the other valley are responsible for 
the polaron dressing of excitons and for splitting the
spectral function into attractive and repulsive exciton-polaron branches.          

As we discuss in details below, renormalization of the exciton binding energy 
and polaronic dressing can be distinguished in experiments. 
We concentrate on the latter by letting $\omega_\mathrm{X}$, $m_\mathrm{X}$, and $D$
be free parameters of the theory, which we allow to have to be weakly dependent on carrier density.
With these assumptions, the polaronic effect on photo-excited excitons can be 
described by the following Hamiltonian
\begin{widetext}
\begin{equation}
\label{HamiltonianModel}
H=\int d\vec{r} \left\{\Psi_\mathrm{e}^+(\vec{r})\left[\frac{\left(\vec{p}_\mathrm{e}+\frac{e}{c}\vec{A}_\mathrm{e}\right)^2}{2m_\mathrm{e}}-\mu_\mathrm{e}\right]\Psi_e(\vec{r}) + \Psi_\mathrm{X}^+(\vec{r})\left[\frac{\vec{p}^2_\mathrm{X}}{2m_\mathrm{X}}+\omega_\mathrm{X}\right]\Psi_\mathrm{X}(\vec{r})-U\; \Psi^+_\mathrm{e}(\vec{r}) \Psi^+_\mathrm{X}(\vec{r}) \Psi_\mathrm{X}(\vec{r}) \Psi_\mathrm{e}(\vec{r}) \right\} 
\end{equation}
\end{widetext}
where $e$, $\mu_\mathrm{e}$, and $m_e$ are the magnitude of charge for electrons ($\hbox{e}$), their chemical potential and mass. $\vec{A}_e=(-By/2,B x/2,0)$ is the vector potential for the magnetic field of strength $B$ in the symmetric gauge.  In Eq.~\ref{HamiltonianModel} $U$ is the short-range attractive exciton-electron interactions and 
$\Psi_e(\vec{r})$ and $\Psi_X(\vec{r})$ are field operators for electrons and excitons in different valleys~\footnote{We consider here only exciton-polarons formed from the ground excitonic state. Excited states and delocalized electron-hole ones are well energy separated from it and are excluded from the consideration. That is why excitons have only the center of mass degree of freedom that is the argument of their annihilation operator $\Psi_\mathrm{X}(\vec{r})$ in Hamiltonian (\ref{HamiltonianModel}) and in wave functions ((\ref{Basis1}) and (\ref{Basis2})).}.  For the sake of definiteness we assume that the excess charge carriers are electrons; the generalization to holes is straightforward.  Note that we have neglected interactions 
between the excess electrons, Since the two valleys can be probed independently by 
using circularly polarized light, we do not introduce a valley index. 

Valley resolved optical conductivity is illustrated schematically by the Feynman diagram depicted in Fig.1-$(\hbox{a})$ and is given by 
\begin{equation}\label{OpticalConductivity}
\sigma(\omega)=\frac{e^2}{h} \, |D|^2 \,  \bar{\epsilon}_\mathrm{X} \, A_\mathrm{X}(\omega,0).
\end{equation}
Here $\bar{\epsilon}_\mathrm{X}=m e^4/\kappa^2 \hbar^2$ is the bare binding energy of excitons in the absence of doping and magnetic field with $\kappa$  to be the effective dielectric constant of the TMDC material. $A_\mathrm{X}(\omega,\vec{p})=-2 \mathrm{Im}[G_\mathrm{X}(\omega,\vec{p})]$
is the spectral function of the excitons.  
In the absence of electrons 
\begin{equation} 
A_\mathrm{X}(\omega,0)=\frac{2\gamma_\mathrm{X}}{\gamma_\mathrm{X}^2+ \Delta \omega^2},
\end{equation} 
where $\gamma_\mathrm{X}$ is a phenomenological exciton radiative lifetime/scattering rate and $\Delta\omega=\omega-\omega_\mathrm{X}$ is the offset frequency 
(In the following we omit the use of $\hbar$ to distinguish wavevector/momentum and frequency/energy, but preserve it elsewhere).  
Our main goal is to calculate $A_\mathrm{X}(\omega,0)$ in the presence of the Fermi sea of excess electrons that 
experience an external magnetic field.  
Before proceeding to the many-body polaronic problem, we briefly review 
some aspects 
of the theories of two-dimensional electrons in a magnetic field (Section III)  
and of two-particle exciton-electron systems (Section IV).

\begin{figure}[t]
	\label{Fig2}
	\vspace{-2 pt}
	\includegraphics[width=7.7 cm]{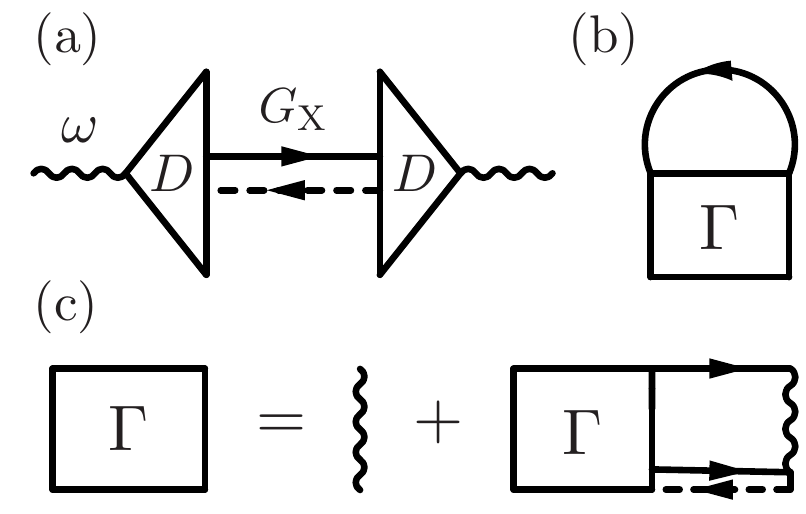}
	\vspace{-2 pt}
	\caption{$\hbox{(a)}$ Schematic representation of Eq.(\ref{OpticalConductivity}), which specifies our 
	approximation for the excitonic contribution to the optical conductivity.  The triangle vertices correspond to the optical matrix elements $D$. The paired solid and dashed lines represent an exciton as a bound state of an 
	electron and a hole and stand for the exciton's free bosonic Green function. $\hbox{(b)}$ 
		Exciton self-energy due to interactions $\Gamma$ with Fermi sea fluctuations. $\hbox{(c)}$ Bethe-Salpeter equation for the exciton/Fermi-sea interaction vertex $\Gamma$. 
		}
\end{figure}
 
\section{III. Landau levels}
In the presence of an external magnetic field, the kinetic-energy spectrum of electrons is quenched to
a set of degenerate Landau levels~\cite{Ezawa,MacDonaldReview}.
In the convenient symmetric gauge, electronic states are labeled by a pair of quantum numbers $|n,l\rangle$. They have 
energy $\epsilon^\mathrm{e}_{n}=\hbar \omega_\mathrm{B}(n+1/2)$ and definite 
orbital momentum $M^\mathrm{z}_{nl}=\hbar (n-l)$, where $\omega_\mathrm{B}=e B/m_e c$ is the Larmor frequency.
Energy does not depend on $l$ and Landau levels have macroscopic 
degeneracy $N_\mathrm{B}=1/2\pi l_\mathrm{B}^2$ per unit area, 
where $l_\mathrm{B}=\sqrt{\hbar c/e B}$ is a magnetic length. When Landau levels are all either completely filled or 
completely empty, electrons form an incompressible quantum system. 
At zero temperature the dependence of their chemical potential $\mu_\mathrm{e}$ on 
concentration $n_\mathrm{e}$ is  discontinuous and is given 
by $\mu_\mathrm{e}=\hbar \omega_\mathrm{B}(1/2+[\nu_\mathrm{e}])$. 
Here $\nu_\mathrm{e}=n_\mathrm{e}/N_\mathrm{B}$ is the total filling factor, 
while $[\nu_\mathrm{e}]$ denotes the integer part of $\nu_\mathrm{e}$ and gives the total number of fully 
occupied Landau levels. 

The states $|nl\rangle$ can be constructed with the help of ladder 
operators $a=l_\mathrm{B}(\Pi_e^x-i \Pi_e^y)/\sqrt{2}\hbar$ and $b=l_\mathrm{B}(K_e^x+i K^y_e)/\sqrt{2}\hbar$ as follows:  
\begin{equation}
\label{LandauStates}
|nl\rangle=\frac{(a^+)^n (b^+)^l}{\sqrt{n!\;l!}\;} |00\rangle,
\end{equation} 
where the ground state is given by
\begin{equation} 
\langle \vec{r}_e|00\rangle= \frac{1}{\sqrt{2\pi l_\mathrm{B}^2}}\exp\left[-\frac{\vec{r}_\mathrm{e}^2}{4 l_\mathrm{B}^2}\right].
\end{equation}
The kinematic, $\vec{\Pi}_\mathrm{e}$, and and magnetic, $\vec{K}_e$, momenta introduced above 
are related to the canonical momentum operator $\vec{p}_\mathrm{e}$ by 
$\vec{\Pi}_\mathrm{e}  = \vec{p}_\mathrm{e}+e \vec{A}_\mathrm{e}/c$, 
and $\vec{K}_e=\vec{p}_e-e \vec{A}_e/c$.
Note that $[\Pi^x_e, \Pi^y_e]= -i \hbar/l_\mathrm{B}$ and $[K^x_e, K^y_e]= i \hbar/l_\mathrm{B}$, so that 
$[a,a^+]=1$, $[b,b^+]=1$, $[a,b]=0$, and $[a,b^+]=0$.  

Our polaronic calculations require manipulations with the
density form-factor of electrons $F_{n' n}^{l' l}(\vec{p})=\langle n' l'| e^{-i \vec{p}\vec{r}_\mathrm{e}}|n l\rangle$. Its explicit form and the derivation of the following helpful identities 
\begin{equation}
\label{DensityFFIdentity1}
\sum_{l_1}\langle F^{l'l_1}_{n'n_1}(-\vec{p}) F^{l_1 l}_{n_1 n}(\vec{p}) \rangle_\vec{p}=\delta_{n' n}^{l' l} \Phi^n_{n_1}\left (\frac{\vec{p}^2 l_B^2}{2}\right),   
\end{equation}
\begin{equation}
\label{DensityFFIdentity2}
\sum_{l' l} F^{l' l}_{n'n}(-\vec{p}') F^{l l'}_{n n'}(\vec{p}) =\delta_{\vec{p}\vec{p}'} N_\mathrm{B}\Phi^n_{n'}\left (\frac{\vec{p}^2 l_B^2}{2}\right),   
\end{equation}
are presented in Appendix A. Here $\langle \cdot\cdot\cdot\rangle_{\vec{p}}$ denotes averaging over direction of momenta $\vec{p}$ and the form-factor $\Phi^n_{n'}(x)$ is given by 
 \begin{equation}
 \label{Phi}
 \Phi^{n'}_n(x)=x^{|n'-n|} \,  \frac{\min[n',n]!}{\max[n',n]!} \, \left\{L^{|n'-n|}_{\min[n',n]}[x]\right\}^2 e^{-x}.
\end{equation}
Here $L^m_n[x]$ is the generalized Laguerre polynomial.

The origin of the macroscopic degeneracy of Landau levels is magnetic translation symmetry.
In the presence of a field, translations are generated by the magnetic momentum $\vec{K}_e$ 
and translational invariance implies that $[H_\mathrm{e},\vec{K}_e]=0$. 
This symmetry is noncommutative since different components of the
magnetic momentum do not commute with each other. 
Physically, the motion of an electron in a magnetic field is along a Larmor orbit, and the components 
of $\vec{K}_\mathrm{e}$ determine its center. The macroscopic degeneracy of Landau levels is 
with respect to the position of the orbit center. 

\section{IV. Two-particle exciton-electron problem} 
In the absence of magnetic field the polaronic exciton-electron problem is simplified by 
the decoupling of the center of mass and relative motion degrees of freedom. 
The separation is guaranteed by translational symmetry. In the presence of magnetic field translational symmetry is replaced by 
noncommutative magnetic 
translation symmetry, defined by $[H,\vec{K}]=0$, and the separation is not possible anymore. 
Here $\vec{K}=\vec{K}_e +\vec{p}_X$ is the \emph{total} magnetic momentum of the
charged electron and the neutral exciton.
The components of $\vec{K}$ do not commute $[K_x, K_y]= i \hbar/l_\mathrm{B}$, and the symmetry implies that $[H,B^+ B]=0$, where $B =l_B(K_x+i K_y)/\sqrt{2}\hbar$ is a new 
ladder operator. The presence of magnetic translation symmetry drastically simplifies the exciton-polaron problem, however its exploitation is not as simple as in the absence of a magnetic field.

For the case of few-particle charged complexes in a 
magnetic field a successful approach has been developed by 
A. Dzyubenko and collaborators~\cite{Dzyubenko1,Dzyubenko2}. It involves a set of Bogoliubov transformations to a basis respecting the symmetry of magnetic translations. The approach requires that all involved particles to be charged and can be applied only for two- and three-particle 
systems. As a result, it cannot be used for the many-body exciton-electron polaronic problem. 
Here we have developed a related approach which also involves a transformation to an
appropriately chosen basis that respects the underlying magnetic translational symmetry, 
but works in a different way.  In this section we present how the transformation works for the two-particle exciton-electron problem.

Before presenting the two-particle basis we employ, it is useful to discuss the more  
obvious direct product representation of the electron plus exciton Hilbert space~\footnote{We consider here only exciton-polarons formed from the ground excitonic state. Excited states and delocalized electron-hole ones are well energy separated from it and are excluded from the consideration. That is why excitons have only the center of mass degree of freedom that is the argument of their annihilation operator $\Psi_\mathrm{X}(\vec{r})$ in Hamiltonian (\ref{HamiltonianModel}) and in wave functions ((\ref{Basis1}) and (\ref{Basis2})).}:
\begin{equation}
\label{Basis1}
	\langle \vec{r}_e \vec{r}_X|n l \vec{p}\rangle=\langle \vec{r}_e |n l\rangle \; e^{i \vec{p} \vec{r}_X}
\end{equation}
These wave functions are normalized over the unit area. For every exciton momentum $\vec{p}$, the product state 
$|n l \vec{p}\rangle$ can be obtained from the corresponding ground 
state $|0 0 \vec{p}\rangle$ with the help of ladder operators $a$ and $b$.
The two-particle Hamiltonian corresponding to (\ref{HamiltonianModel}) in this basis can be decomposed as follows 
\begin{widetext}
\begin{equation}\label{ME1}
\begin{split}
H=\sum_{nl \vec{p}} \left(\xi^\mathrm{e}_n+\frac{\vec{p}^2}{2 m_\mathrm{X}}+\omega_{X}\right)|nl\vec{p}\rangle\langle n l\vec{p}| - \sum_{nl \vec{p}} \sum_{n'l' \vec{p}'} U F_{n' n}^{l' l}(\vec{p}'-\vec{p})  |n' l' \vec{p}'\rangle\langle n l\vec{p}|.
\end{split}
\end{equation}
\end{widetext}
Here $\xi^\mathrm{e}_n=\hbar \omega_\mathrm{B}(n+1/2)-\mu_\mathrm{e}$ is the electron energy and $F_{n' n}^{l' l}(\vec{p})=\langle n' l'| e^{-i \vec{p}\vec{r}_\mathrm{e}}|n l\rangle$ is the density form-factor. The product basis is the one naturally employed in diagrammatic perturbation theory 
because it is the representation in which the kinetic energy, and hence the 
bare electron and exciton propagators, are diagonal.  
On the other hand, the form-factor $F_{n' n}^{l' l}(\vec{p}-\vec{p}')$ couples both Landau levels $n$ and orbits 
$l$ in a complicated manner. The simplifications afforded by magnetic translation symmetry 
are not explicit. 

Now we construct a new basis. 
We notice that the magnetic momentum of electrons $\vec{K}_\mathrm{e}$, respected in the usual basis, can be transformed to the total one $\vec{K}$ as $U \vec{K}_\mathrm{e}U^+=\vec{K}$ with the unitary operator $U=e^{-i \vec{p}_\mathrm{X} \vec{r}_\mathrm{e}}$. This observation 
motivates the introduction of a new basis as 
follows:
\begin{equation}
\label{Basis2}
\langle \vec{r}_e\vec{r}_X|\overline{n l \vec{p}}\rangle \equiv \langle \vec{r}_e\vec{r}_X| U |n l \vec{p}\rangle = 
\langle \vec{r}_e |nl\rangle \; e^{i \vec{p} (\vec{r}_X-\vec{r}_e)}.
\end{equation}
The new states, which will be distinguished by an overbar accent,
are labeled by the same quantum numbers, $n$ and $l$ and $\vec{p}$, as the product basis. 
Importantly, excited states  $|\overline{n l \vec{p}}\rangle$ for each momentum $\vec{p}$ can be constructed from the corresponding ground state $|\overline{0 0 \vec{p}}\rangle$ with the help of $A=U a U^+=l_B(\Pi^x-i \Pi^y)/\sqrt{2}\hbar$ and $B=U b U^+=l_B(K^x+i K^y)/\sqrt{2}\hbar$. Here $\vec{\Pi}=\vec{\Pi}_e+\vec{p}_X$ and $\vec{K}=\vec{K}_e+\vec{p}_X$ are \emph{total} kinematic and magnetic momenta we introduced above. As a result, new states (\ref{Basis2}) are eigenstates of $B^+B$ as $B^+B |\overline{nl\vec{p}}\rangle=l |\overline{n l \vec{p}}\rangle$ and magnetic translation symmetry is respected. The overlap between the basis states is the density 
form-factor $\langle n' l' \vec{p}'|\overline{n l \vec{p}}\rangle=\delta_{\vec{p}' \vec{p}}F^{l'l }_{n'n}(\vec{p})$, and the transformation between the two representations can be alternatively presented as
\begin{equation}
|\overline{nl\vec{p}}\rangle=\sum_{n' l'} F^{l' l}_{n' n}(\vec{p}) |n'l'\vec{p}\rangle. 
\end{equation}
In the new representation, the Hamiltonian of two-particle exciton-electron problem can be decomposed as
\begin{widetext}
	\begin{equation*}\label{ME2}
	\begin{split}
	H=\sum_{nl \vec{p}}\left\{ \left(\xi^\mathrm{e}_n+\frac{\vec{p}^2}{2 \mu_\mathrm{T}}+\omega_{X}\right)|\overline{nl\vec{p}}\rangle\,\langle \overline{n l\vec{p}}| - \frac{n \hbar p^+}{\sqrt{2} m_\mathrm{e}} |\overline{n-1,l\vec{p}}\rangle\langle \overline{n l\vec{p}}| - \frac{n \hbar p^-}{\sqrt{2} m_\mathrm{e}} |\overline{n,l\vec{p}}\rangle\langle\overline{n-1, l\vec{p}}|  \right\}  - \sum_{nl \vec{p} \vec{p}'} U  |\overline{n l \vec{p}'}\rangle\langle \overline{n l\vec{p}}|.
	\end{split}
	\end{equation*}
\end{widetext}
Here we introduced compact notations $p^\pm=p_\mathrm{x}\pm i p_\mathrm{y}$, and $\mu_\mathrm{T}=m_\mathrm{e} m_\mathrm{X}/(m_\mathrm{e}+m_\mathrm{X})$ is the reduced mass of exciton and electron. The main advantage of new basis is that Landau orbits with different intra-Landau-level 
label, $l$, are decoupled from each other. 
Moreover, interaction matrix elements between the electron and the exciton are also diagonal
with respect to the Landau level kinetic energy index, $n$.  As we see below, these properties make the basis convenient for analysis of the exciton-electron scattering problem.

\section{V. Exciton-polaron problem}

We now return to the many-body exciton-electron problem. 
The approximation we employ for the dressed exciton propagator is
illustrated diagrammatically in Fig.2-$\hbox{b}$ and $\hbox{c}$. 
The many-body scattering amplitude $\Gamma^{l'l}_{n'n}(\omega,\vec{p}',\vec{p})$  
is used to construct the exciton self-energy, illustrated in Fig.2-$\hbox{b}$, that describes the 
the polaronic dressing.  $\Gamma^{l'l}_{n'n}(\omega,\vec{p}',\vec{p})$ accounts for the scattering 
amplitude between the exciton-electron two-particle product states
$|nl\vec{p}\rangle$ and $|n'l'\vec{p}'\rangle$ via repeated electron-exciton interactions 
in a ladder diagram approximation. 
After summation over Matsubara frequencies and analytical continuation, the
Bethe-Salpeter equation for the $\Gamma-$vertex is, 
\begin{widetext}
	\begin{equation}
	\label{BS1}
	\Gamma^{l'l}_{n'n}(\omega,\vec{p}',\vec{p})=-U F^{l'l}_{n'n}(\vec{p}'-\vec{p})-
	\sum_{n_1 l_1} \sum_{\vec{p}_1} U F^{l'l_1}_{n'n_1}(\vec{p}'-\vec{p}_1)  \frac{ \bar{\nu}^\mathrm{e}_{n_1}}{\Delta\omega-\xi^\mathrm{e}_{n_1}-\epsilon^\mathrm{X}_{\vec{p}_1}+i \gamma_\mathrm{X}} \Gamma^{l_1 l}_{n_1n}(\omega,\vec{p}_1,\vec{p}).
	\end{equation}
\end{widetext}
Here $\epsilon_\vec{p}^\mathrm{X}=\vec{p}^2/2m_\mathrm{X}$ is the energy of excitons. The factor $\bar{\nu}^\mathrm{e}_n=1-\nu^\mathrm{e}_n$ accounts for Pauli blocking by excluding filled states from the scattering, while $\nu^\mathrm{e}_n$ is the filling factor of Landau level $n$. Because magnetic translation symmetry is not respected in this usual basis, all two-particle states 
$|nl\vec{p}\rangle$ and  $|n'l'\vec{p}'\rangle$ are coupled with each other in a complicated way that makes 
further manipulations problematic.

Now we transform the scattering amplitude $\Gamma$ to the more 
convenient basis $\bar{\Gamma}$ using: 
\begin{equation*}
\Gamma^{l' l}_{n' n}(\omega, \vec{p}', \vec{p})=\sum_{\begin{smallmatrix}
	l_1 l_2 \\ n_1 n_2
	\end{smallmatrix}} F^{l' l_1}_{n' n_1} (\vec{p}') \bar{\Gamma}^{l_1 l_2}_{n_1 n_2}(\omega, \vec{p}',\vec{p}) F^{l_2 l}_{n_2 n} (-\vec{p}). 
\end{equation*}
Transformation of
Eq. (\ref{BS1}) to this representation leads to 
\begin{widetext}
	\begin{equation}
	\label{BS2}
	\bar{\Gamma}^{l'l}_{n'n}(\omega,\vec{p}',\vec{p})=-U \delta^{l'l}_{n'n}-
	\sum_{n_1 l_1} \sum_{n_2 l_2} \sum_{\vec{p}_1} U F^{l'l_1}_{n'n_1}(-\vec{p}_1)  \frac{ \bar{\nu}^\mathrm{e}_{n_1}}{\Delta\omega-\xi^\mathrm{e}_{n_1}-\epsilon^\mathrm{X}_{\vec{p}_1}+i\gamma_\mathrm{X}} F^{l_1 l_2}_{n_1 n_2}(\vec{p}_1) \bar{\Gamma}^{l_2 l}_{n_2 n}(\omega,\vec{p}_1,\vec{p}).
	\end{equation}
\end{widetext}
Because the interaction matrix elements
in the new basis do not depend on transferred momentum $\vec{p}'-\vec{p}$,  the scattering amplitude is 
momentum independent $\bar{\Gamma}_{n' n}^{l' l}(\omega)$.  Summation over $l_1$ and averaging over the direction of momentum $\vec{p}_1$ with the help of the identity (\ref{DensityFFIdentity1}) decouples all scattering channels $\bar{\Gamma}_{n' n}^{l' l}(\omega)=\delta^{l'l}_{n'n} \bar{\Gamma}_{n' n}(\omega)$ and makes the Bethe-Saltpeter equation algebraic $\bar{\Gamma}_n(\omega)=-U-U  \Pi_n(\omega)\bar{\Gamma}_n(\omega)$. Here the kernel $\Pi_n(\omega)$ is given by 
\begin{equation}
\label{Pi}
\Pi_{n'}(\omega)=\sum_{\vec{p} n}\frac{ \bar{\nu}^\mathrm{e}_{n} \Phi^{n'}_{n} \left(\frac{\vec{p}^2 l_B^2}{2}\right) }{\Delta\omega-\xi^\mathrm{e}_{n}-\epsilon^\mathrm{X}_{\vec{p}}+i \gamma_\mathrm{X}},
\end{equation}
and $\Phi^{n'}_n(x)$ is the form-factor introduced in (\ref{Phi}). The kernel $\Pi_n(\omega)$ has the ultraviolet logarithmic divergence that is also  
present in the absence of a magnetic field~\cite{EfimkinMacDonald}. It is instructive to introduce the energy cutoff $\Lambda$ and rewrite the scattering amplitude $\bar{\Gamma}_n$ as follows  
\begin{equation}
\label{Gamma}
\bar{\Gamma}_n=-\frac{1}{N_\mathrm{T}\ln[\frac{\Lambda}{\epsilon_\mathrm{T}}]+\Pi_n(\omega,\Lambda)}.
\end{equation}
Here $\epsilon_\mathrm{T}=\Lambda\exp[-1/N_\mathrm{T} U]$ is the binding energy of the two-body state of electron and exciton, that is the simplified model of trion,  in the absence of magnetic field, while  $N_\mathrm{T}=\mu_\mathrm{T}/2 \pi \hbar^2$ is their reduced density of states. Because 
$\bar{\Gamma}_n$ does not depend explicitly on $\Lambda$ and $U$, we treat $\epsilon_\mathrm{T}$ as an independent input parameter~\footnote{The logarithmic ultraviolet divergence of the kernel $\Pi_n(\omega)$, given by (\ref{Pi}), comes from Landau levels with high kinetic energy index $n$. In Appendix C we use the momentum space coarsening picture, that is valid for their formation, to show to that their contribution can be approximated by the related expressions derived by us in the absence of magnetic field. In the latter case, briefly overviewed in Appendix B, $\Gamma$-vertex does not depend explicitly on $\Lambda$ and $U$, but only on their combination $\epsilon_\mathrm{T}$.}. 

The many-body scattering amplitude $\bar{\Gamma}_n$ defines the self-energy of excitons $\Sigma(\omega,\vec{p})$ depicted in Fig.1-b. After summation over Matsubara frequencies and analytical continuation, the self-energy is,
\begin{equation}
\Sigma^\mathrm{X}_{\vec{p}' \vec{p}}(\omega)=\sum_{n l}\nu^\mathrm{e}_n \Gamma^{l l}_{n n} (\omega+\xi^\mathrm{e}_n, \vec{p}', \vec{p}),
\end{equation} 
where $\Gamma$ is expressed in the product state representation.
To take advantage of translational symmetry we make a transformation to the modified 
representation:

\begin{equation}
\Sigma^\mathrm{X}_{\vec{p}' \vec{p}}(\omega)=\sum_{\begin{smallmatrix}
	n l \\ n' l'
	\end{smallmatrix}}\nu^\mathrm{e}_n F^{l l'}_{n n'}(\vec{p}') \bar{\Gamma} _{n'} (\omega+\xi^\mathrm{e}_n) F^{l' l}_{n' n}(-\vec{p})
\end{equation}
Summation over Landau orbits $l$ and $l'$ with the help of identity (\ref{DensityFFIdentity2}) yields 
a self-energy that is diagonal in momentum space given by:
\begin{equation}
\Sigma_\mathrm{X}(\omega,\vec{p})=N_\mathrm{B}\sum_{n n'} \nu^\mathrm{e}_n \Phi^n_{n'}\left(\frac{\vec{p}^2 l_\mathrm{B}^2}{2}\right)  \bar{\Gamma} _{n'} (\omega+\xi^\mathrm{e}_n).
\end{equation}
Here $N_\mathrm{B}$ is the Landau level
degeneracy factor and $\Phi^{n'}_n(x)$ is the form-factor introduced in (\ref{Phi}). 
Only excitons with zero momentum $\vec{p}=0$ are optically active and the self-energy expression reduces further to,  
\begin{equation}
\label{Sigma}
\Sigma_\mathrm{X}(\omega,0)=N_\mathrm{B}\sum_{n} \nu^\mathrm{e}_n \bar{\Gamma}_{n} (\omega+\xi^\mathrm{e}_n).
\end{equation}
The self-energy modifies the spectral function of excitons $A_\mathrm{X}(\omega,0)=[\Delta\omega-\Sigma_\mathrm{X}(\omega,0)+i \gamma_\mathrm{X}]^{-1}$ that is proportional to the optical conductivity, given by Eq.~(\ref{OpticalConductivity}) and probed in absorption experiments. 

The closed set of equations (\ref{Pi}), (\ref{Gamma}) and (\ref{Sigma}) describes the polaronic dressing of excitons.  In the limit of vanishing magnetic field $\hbar \omega_\mathrm{B}\ll\epsilon_\mathrm{T}$ these equations reduce to the ones derived in our previous work~\cite{EfimkinMacDonald} which are briefly summarized in Appendix B. 
A formal proof of this statement, constructed by 
employing the momentum space coarsening picture of Landau level
formation, is presented in Appendix C. 
In the opposite regime of strong magnetic fields $\epsilon_\mathrm{T}\ll\hbar \omega_\mathrm{B}$
the assumption of our polaronic theory $\hbar\omega_\mathrm{B}\ll \epsilon_\mathrm{X}$ are not well satisfied.
In this regime, which we discuss in Appendix D, only a few Landau levels are relevant.

\section{V. Results}
For $T\ll \epsilon_\mathrm{T}$, the exciton 
spectral function depends on four dimensionless control parameters: offset frequency $\Delta\omega/\epsilon_\mathrm{T}$, magnetic field $\hbar \omega_\mathrm{B}/\epsilon_\mathrm{T}$, exciton decay/scattering rate $\gamma_\mathrm{X}/\epsilon_\mathrm{T}$ and doping $\epsilon_\mathrm{F}/\epsilon_\mathrm{T}$. Here $\epsilon_\mathrm{F}$ is the Fermi energy of electrons in the absence of magnetic field
which is related to the total filling factor by
$\nu_\mathrm{e}=n_e/N_\mathrm{B}=\epsilon_\mathrm{F}/\hbar \omega_\mathrm{B}$. 
We fix $\gamma_\mathrm{X}/\epsilon_\mathrm{T}=0.1$ and consider two 
values for the dimensionless magnetic field
$\hbar\omega_\mathrm{B}/\epsilon_\mathrm{T}=0.5;\; 1$, which we refer to as to
the weak, and moderate magnetic field cases. In TMDC materials the band masses for conduction and valence bands are almost the same and we used $m_\mathrm{X}=2m_\mathrm{e}$ in numerical calculations.

In Fig.~3~(a) we present our results for the exciton spectral function $A_\mathrm{X}(\omega,0)$ 
in the absence of a magnetic field. Due to interactions with the Fermi sea of electrons, the excitonic feature splits into a 
redshifted attractive exciton-polaron branch, normally identified as a trion branch, and
a bluesifted repulsive exciton-polaron branch, normally identified as an exciton branch. These
branches are the many-body generalizations of the zero center of mass momentum  
bound and unbound exciton-electron states with and smoothly evolve from them. 
With increasing doping the attractive branch peak evolves between two approximately linear 
behaviors as a function of Fermi energy $\epsilon_\mathrm{T}^*=-\epsilon_\mathrm{T}-\epsilon_\mathrm{F}$ and  $\epsilon_\mathrm{T}^*=-3\epsilon_\mathrm{T}/2-\epsilon_\mathrm{F}/2$, 
whereas the repulsive branch peak varies linearly as $\epsilon_\mathrm{X}^*=\epsilon_\mathrm{F}/2$. 
These approximate dependences on carrier-density are represented in Fig.~3 by dashed lines. 

The exciton spectral function $A_\mathrm{X}(\omega,0)$  is plotted in Figs.~3~(b-d)
at five different values of magnetic field. 
The attractive exciton-polaron branch is only weakly influenced by the magnetic field.  For $T=0$, the 
spectral function has weak cusps at integer filling factors because of the singular dependence on field 
of the filling factors of particular Landau levels when they are first occupied and when they are empty.  
The attractive
mode is the many-body generalization of the bound state of an exciton and an 
electron and is not strongly 
sensitive to the discretization of the 
electronic state energies due to Landau level formation. 
Even for the magnetic field $\hbar \omega_\mathrm{B}\sim \epsilon_\mathrm{T}$ 
the attractive branch can still be very well approximated by the 
curve $\epsilon_\mathrm{T}^*=-3\epsilon_\mathrm{T}/2-\epsilon_\mathrm{F}/2$.

\begin{figure}[t]
	\label{Fig4}
	\vspace{-2 pt}
	\includegraphics[width=7.9 cm]{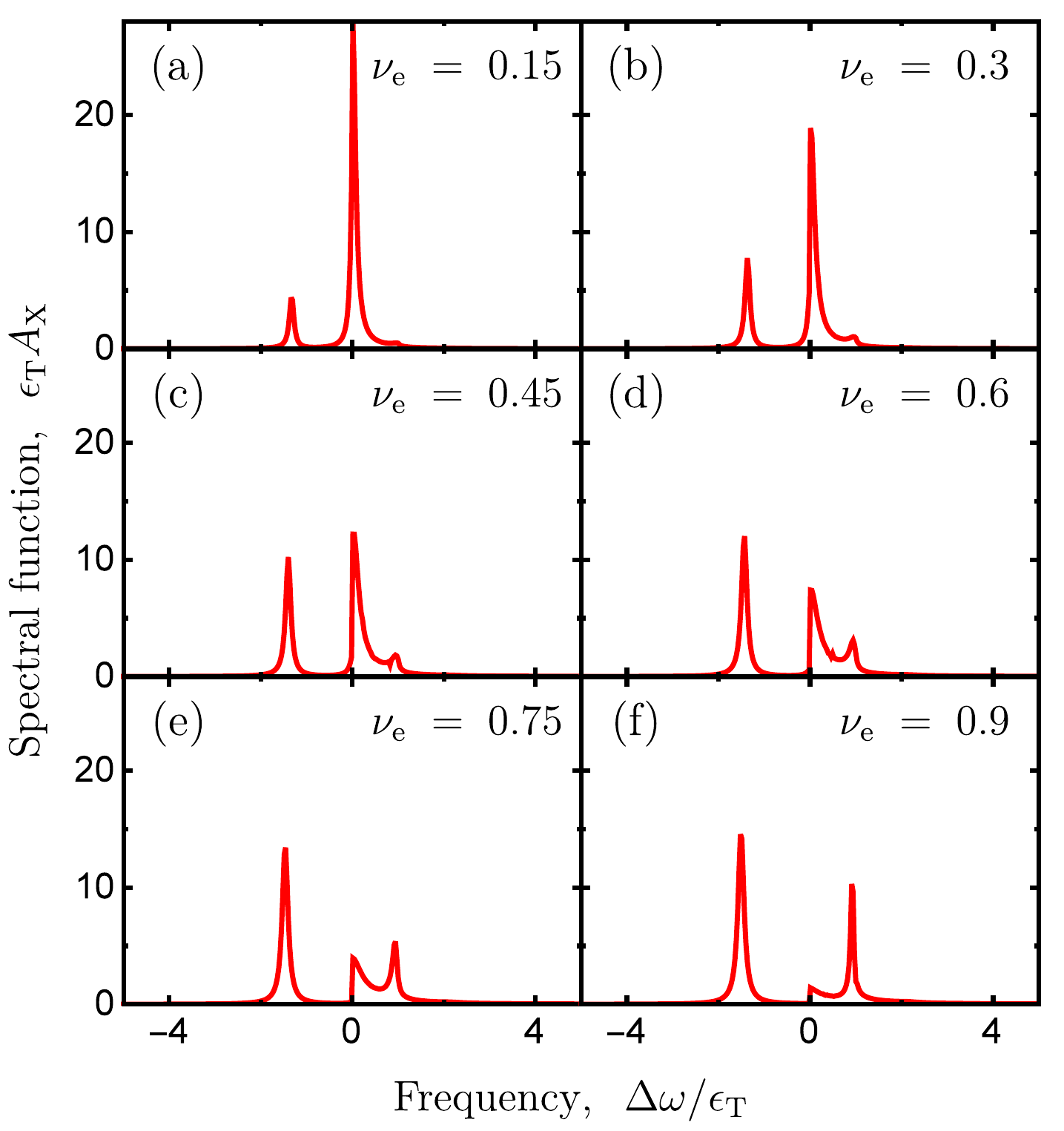}
	\vspace{-2 pt}
	\caption{Frequency dependence of the excitonic spectral function $A_\mathrm{X}(\omega,\vec{p}=0)$ at different fillings factors ranging from $\nu_\mathrm{e}=0.15$ ($\hbox{a}$) to $\nu_\mathrm{e}=0.9$ ($\hbox{f}$). The strength of the magnetic field is  $\hbar\omega_\mathrm{B}/\epsilon_\mathrm{T}=1$.  The partial weight of the repulsive 
	exciton-polaron branch 
	near $\Delta\omega=0$ disappears when
	Landau level $n=0$ becomes fully occupied at $\nu_\mathrm{e}=1$ where the weight of the first 
	exciton-cyclotron resonance at $\Delta \omega=\hbar \omega_\mathrm{B}$ achieves its maximum.} 
\end{figure} 

As seen in Figs.~3, we find that 
the repulsive exciton-polaron branch is more strongly influenced by the magnetic field and 
experiences oscillatory behavior. 
It is the many-body generalization of the unbound exciton-electron state which experiences
Landau quantization. 
The main dispersionless exciton-polaron feature near $\Delta\omega=0$ originates from
exciton creation processes that are not accompanied by inter-Landau level 
electronic transitions. The set of blueshifted peaks separated by the 
energy $\hbar\omega_\mathrm{B}$ can be attributed to combined exciton-cyclotron resonances,
in which excitons are created and their dressing results in electronic transitions to unoccupied 
higher Landau levels.  
Our approximations involve the summation of all ladder exciton-electron scattering diagrams, nevertheless the doping dependence of these features can be  explained based on the second-order perturbation theory, which we present in Appendix C. In the perturbation theory the contribution of intra-level transitions, responsible for the feature at $\Delta \omega=0$, is proportional to $\nu_\mathrm{n}^\mathrm{e} \bar{\nu}_\mathrm{n}^\mathrm{e}$.  
Due to the Pauli blocking effect the feature smoothly disappears at integer filling factors and achieves maximum when the corresponding Landau level is half-filled. The contribution of electronic transitions between different Landau levels $n$ and $n'$, corresponding to the exciton-cyclotron resonance at $\Delta \omega=\hbar \omega_\mathrm{B}(n'-n)$, is proportional to $\nu_n^\mathrm{e} \bar{\nu}_{n'}^\mathrm{e}$. As a result, it achieves a maximum when Landau level $n$ is fully filled and level $n'$ is fully empty. 
This doping dependence of the spectral function $A_\mathrm{X}(\omega,0)$ is 
captured in 
Fig.~4~($\hbox{a}$-$\hbox{f}$) for the case of the strong magnetic field 
$\hbar \omega_\mathrm{B}/\epsilon_\mathrm{T}=2$ and the filling factor range from $\nu_\mathrm{e}=0.15$  to $\nu_\mathrm{e}=0.9$.

With decreasing magnetic field the number of additional blueshifted exciton-cyclotron resonances grows dramatically. 
They compete with each other and finally merge into the broad dispersive resonance representing repulsive exciton-polaron branch. 

\section{VI. Discussion and Conclusions}
We have investigated the polaronic dressing of excitons by excess electrons
using a partially phenomenological approach that treats
the exciton creation energy $\omega_\mathrm{X}$, the exciton mass $m_\mathrm{X}$, and the optical 
matrix element for photon to exciton coversion $D$, as parameters. 
The presence of excess electrons makes these parameters doping dependent. 
However, the only role of $\epsilon_{X}$ is to shift the absorption spectrum rigidly in frequency,
and the only role of $D$ is to rigidly scale the absorption strength, which is proportional to $|D|^2$.
The goal of our theory is to explain the frequency-splitting between different features, and 
their relative amplitudes/spectral weight, 
which are independent of $D$ and $\omega_\mathrm{X}$. 
The dependence of absorption on excitons mass is much more complicated, but 
fortunately we believe that exciton mass renormalization is weak.
Our earlier zero-magnetic-field calculations~\cite{EfimkinMacDonald}, which combined static 
RPA screening with the Hartree-Fock approximation, suggest that 
the carrier-density dependent change in exciton mass 
$\Delta m\lesssim  0.07 m_\mathrm{e}$.  
To the best of our knowledge measurements of the renormalization 
of electron and hole 
masses in TMDCs, which could substantiate these estimates, 
have not yet been reported.

In TMDC monolayers the trion binding energy $\epsilon_\mathrm{T}\approx 30 \;\hbox{meV}$,
implying that the exciton-polaron regime $\epsilon_\mathrm{F}\lesssim \epsilon_\mathrm{T}$ 
is active at carrier densities smaller than $n_\mathrm{e}\lesssim 5 \times  10^{12} \; \hbox{cm}^{-2}$.  At 
carrier densities $n_\mathrm{e} \lesssim 10^{12} \; \hbox{cm}^{-2}$, the influences of carriers can be 
safely estimated from the three-particle trion physics.  At carrier densities 
$n_\mathrm{e}\gtrsim 10^{13} \; \hbox{cm}^{-2}$, the Fermi energy of excess electrons becomes 
comparable even to the exciton binding energy which is renormalized to substantially smaller values, 
absorption is then  governed by the interplay of 
Mahan and Mott exciton physics~\cite{TrionExperiment1,ExcIon,TMDCExcitonReview},
which has been extensively studied in semiconductor quantum wells~\cite{QWMF1,QWMF2,QWMF3,QWMF4,QWMF5,QWMF6}. The zero temperature limit we used for numerical calculations implies $T\ll \epsilon_\mathrm{T}$~\cite{PolaronFiniteT}, that corresponds to wide temperature range $T\ll 330 \; \hbox{K}$. 
The criterion 
$\hbar \omega_\mathrm{B}\sim \epsilon_\mathrm{T}\approx\; 85 \; \mathrm{T}$
can be used to identify the scale of magnetic field that is required to strongly 
alter the polaronic dressing of excitons. The magnetic oscillations we have predicted 
for the repulsive exciton-polaron branch can be observed at smaller magnetic 
fields if $\gamma_\mathrm{X}\sim T \ll \hbar \omega_\mathrm{B}$. 
For radiative lifetime/scattering rate $\gamma_\mathrm{X}=5\;\hbox{meV}$ reported in doped TMDC samples~\cite{TrionExperiment7} and temperature $T\approx 50\;\hbox{K}$, this condition is well satisfied for 
the magnetic field $B\approx 65\; \hbox{T}$ employed in recent experiments~\cite{TMDCStrongMF}. 
	
The optical properties of the 
conventional semiconductor quantum well systems $\hbox{GaAs}$ and $\hbox{CdTe}$ in a magnetic field 
have been extensively studied~\cite{QWMF1,QWMF2,QWMF3,QWMF4,QWMF5,QWMF6} 
(See \cite{WQReview} for a review). For $\hbox{CdTe}$ quantum wells with trion binding 
energy $\epsilon_\mathrm{T}=2.1\,\hbox{meV}$, the exciton-polaron regime 
$\epsilon_\mathrm{F}\sim\epsilon_\mathrm{T}$ corresponds $n_\mathrm{e}\sim10^{11}\;\hbox{cm}^{-2}$, and
the magnetic field required to influence the polaronic dressing is equal to $B\sim 2~\hbox{T}$. 
Robust attractive exciton-polarons, usually identified as singlet trions, and
combined exciton-cyclotron resonance peaks
that are blue-shifted from the repuslive exciton-polarons, usually identified as excitons, 
have both been reported. 
The resonances have been explained by using second order perturbation theory to account for 
Coulomb interactions~\cite{ShakeUp2,ShakeUpTheory1,ShakeUpTheory2}, as we briefly
discuss in Appendix D. It should be noted that combined exciton-cyclotron resonances 
have been observed only in clean samples and even then only at low temperatures~\cite{ShakeUp1,ShakeUp2}. 
In our theory the crossover to the zero-field repulsive-polaron absorption features has a complex oscillatory behavior which, to the best of our knowledge, has not yet been observed. We expect that the 
improving quality of two-dimensional materials will make it possible to resolve the complex 
magnetic-field dependent structure of the repulsive-polaron absorption feature in the future. 

At enough strong magnetic fields the absorption spectra of the conventional 
semiconductor $\hbox{GaAs}$ and $\hbox{CdTe}$ quantum wells 
acquires an additional feature between the
attractive and repulsive exciton-polaron branches~\cite{WQReview}, which is 
attributed triplet trions. Whereas exchange interactions limit the formation of triplet trions 
in the absence of a magnetic field, its has shown that the interplay of the Zeeman effect, kinetic 
energy quenching, and exchange interactions can favor their formation at finite magnetic field~\cite{MacDonaldTrion,Dzyubenko1,Dzyubenko2,TrionQHTheory1,TrionQHTheory2}. 
The trion scenario is valid at very low densities $\epsilon_\mathrm{F}\ll\epsilon_\mathrm{T}$, while the 
interplay at elevated doping $\epsilon_\mathrm{F}\sim\epsilon_\mathrm{T}$ in the polaronic regime is still 
not well understood. Singlet and triplet trions in conventional semiconductors correspond 
partially to inter-valley and intra-valley excitions in TMDC materials. 
The physics of intra-valley exciton-polarons cannot be specifically 
addressed using our approach, and its consideration is postponed to future work.

We have neglected correlations between excess electrons. The latter are likely to be especially  
important and interesting in the fractional quantum Hall effect regime.
For example we can anticipate that trion binding energies will jump upon passing from 
below to above a filling factor at which an incompressible state appears. 
Recently, additional blueshifted peaks have been observed at some special filling factors ~\cite{FQHE1,FQHE2}.  
We have postponed an attempt to account for the role of Coulomb correlations in the integer and fractional quantum Hall regimes to future work.

To conclude, we have developed an approach for the many-body exciton-polaron problem in magnetic field. The approach involves a transformation to a basis that respects the noncommutative symmetry of magnetic translations.  We have shown that the attractive exciton-polaron branch is very weakly influenced by the magnetic field.
The repulsive branch is strongly modified by the magnetic field, however, and 
exhibits oscillatory behavior and spectral splitting related to Landau quantization. 
The additional peaks correspond to combined exciton-cyclotron resonances and we expect
that their future experimental study will be revealing.  

This material is based upon work supported by the Army Research Office under Award 
No. W911NF-15-1-0466 and by the Welch Foundation under Grant No. F1473.   
The authors acknowledge helpful interactions with Atac Imamoglu, Fengcheng Wu, 
Scott Crooker and Jesper Levinsen. 

 \bibliography{QHEPolaronBIB}
 
 \section{A. The density form-factor}\label{Appendix A}
 Here we present a set of useful relations for the density form-factor 
 $F_{n' n}^{l' l}(\vec{p})=\langle n' l'| e^{-i \vec{p}\vec{r}_\mathrm{e}}|n l\rangle$ that is given by  
 \begin{equation}
 \label{F}
 F_{n' n}^{l' l}(\vec{p})=e^{-\frac{\vec{p}^2 l_B^2}{2}}G^{n'}_{n}(p^-) G^{l'}_ {l}(p^+).
 \end{equation}
 Here we have introduced the compact notation
 $p^\pm=p^x\pm i p^y$, the function $G^{n'}_n (p^\pm)$ is given by
 \begin{equation*}
 \label{G}
 G^{n'}_{n}(p^\pm)=\begin{cases}
 \left(\frac{p^\pm l_B}{i^2\sqrt{2}}\right)^{n'-n}\sqrt{\frac{n!}{n'!}}\;  L_{n}^{n'-n}\left[\frac{\vec{p}^2 l_B^2}{2}\right], \;\;\;\;n'\ge n\\ \left(\frac{p^\mp l_B}{\sqrt{2}}\right)^{n-n'}\sqrt{\frac{n'!}{n!}} \; L_{n'}^{n-n'}\left[\frac{\vec{p}^2 l_B^2}{2}\right], \;\;\;\;n'<n
 \end{cases}
 \end{equation*}
 and $L^m_n[x]$ is the generalized Laguerre polynomial.  
 The polaronic calculations described in the main text
 employ a number of useful identities~\cite{MacDonaldReview} involving $G^{n_2}_{n_1}$:
 \begin{align}
 \label{G1}
 G^{n'}_{ n} (0) = \delta_{n' n},\\
 \label{G2}
 \sum_{n} G^n_n (p^\pm) = (2\pi)^2 N_\mathrm{B}\delta(\vec{p})\equiv N_\mathrm{B} \delta_{\vec{p},0},\\
 \label{G3}
 \sum_{n''} e^{-\frac{p_2^\mp p_1^\mp l_\mathrm{B}^2}{2}} G^{n'}_{n''} (p_2^\pm) G^{n''}_{n}(-p_1^\pm)= G^{n'}_{n}(p_2^\pm-p_1^\pm).
 \end{align}
 The relation for the products of form-factors we use in the main part of the paper:
 \begin{widetext}
 \begin{equation*}
 \sum_{l_1}\langle F^{l'l_1}_{n'n_1}(-\vec{p}) F^{l_1 l}_{n_1 n}(\vec{p}) \rangle_\vec{p}= \delta_{l' l} e^{-\frac{\vec{p}^2 l_B^2}{2}} \langle G^{n'}_{n_1}(p^-) G^{n_1}_{n}(-p^-)\rangle_\vec{p}=  \delta_{n' n}^{l' l}  e^{-\frac{\vec{p}^2 l_B^2}{2}} G^{n}_{n_1}(p^-) G^{n_1}_{n}(-p^-)\equiv
  \delta_{n' n}^{l' l} \Phi^n_{n_1}\left (\frac{\vec{p}^2 l_B^2}{2}\right), 
 \end{equation*}
 and
  \begin{equation*}
  \sum_{l' l }\langle F^{l'l}_{n'n}(-\vec{p}') F^{l l'}_{n n'}(\vec{p}) \rangle_\vec{p}=  
  \delta_{\vec{p}' \vec{p}} N_\mathrm{B} e^{-\frac{\vec{p}^2 l_B^2}{2}}  G^{n'}_{n}(p^-) G^{n}_{n'}(-p^-) \equiv
  \delta_{\vec{p}' \vec{p}} N_\mathrm{B} \; \Phi^n_{n_1}\left (\frac{\vec{p}^2 l_B^2}{2}\right),
  \end{equation*}
 \end{widetext}
were derived using these identies.
Here $\langle \cdot\cdot\cdot\rangle_{\vec{p}}$ denotes an average
 over the direction of the momentum $\vec{p}$ and $\Phi^n_{n'}(x)$ is given by 
 \begin{equation}
 \label{PhiA}
 \Phi^{n'}_n(x)=x^{|n'-n|} \,  \frac{\min[n',n]!}{\max[n',n]!} \, \left\{L^{|n'-n|}_{\min[n',n]}[x]\right\}^2 e^{-x},
 \end{equation}
 where $L^m_n[x]$ is the generalized Laguerre polynomial.

 \section{B. Exciton-polaron dressing in the absence of a magnetic field}
Here we briefly present the main steps for the exciton dressing into exciton-polarons without  magnetic field. Without magnetic field the total momentum of exciton and electron $\vec{P}=\vec{p}_\mathrm{e}+\vec{p}_\mathrm{X}$ is a good quantum number and the $\Gamma$-vertex $\Gamma(\omega,\vec{P},\vec{p'},\vec{p})$ denotes the amplitude of scattering between relative momenta $\vec{p}'$ and $\vec{p}$. For contact interactions between electrons and excitons it becomes independent of $\vec{p}'$ and $\vec{p}$ and the Bethe-Salpeter equation reduces to the algebraic one $\Gamma(\omega,\vec{P})=-U-U\Pi(\omega,\vec{P})\Gamma(\omega,\vec{P})$. Here the kernel $\Pi(\omega,\vec{P})$ is given by 
\begin{equation}
\label{Pi0}
\Pi(\omega,\vec{P})=\sum_\vec{p}\frac{1-n_\mathrm{F}(\epsilon_{\vec{p}+\vec{P}/3})}{\omega^+-\epsilon_\mathrm{X}-\frac{\vec{P}^2}{2M_\mathrm{T}}- \frac{\vec{p}^2}{2\mu_\mathrm{T}}+\epsilon_\mathrm{F}}.
\end{equation}
Here $M_\mathrm{T}=3m_\mathrm{e}$ and $\mu_\mathrm{T}=2 m_\mathrm{e}/3$ are the total and reduced masses of the exciton-electron system in the considered case $m_\mathrm{X}=2 m_\mathrm{e}$. Direct calculations result in
\begin{equation}
\label{Gamma0}
\Gamma(\omega,\vec{P})=\frac{2\pi \hbar^2}{\mu_\mathrm{T}}\frac{1}{\log\left[\frac{\epsilon_\mathrm{T}}{\Omega}\right]+\bm{i} \pi},
\end{equation}
 where $\epsilon_\mathrm{T}=\Lambda\exp[-1/N_\mathrm{T} U]$ is
 the binding energy of the two-body state of electron and exciton, and $\Lambda$ is the ultraviolet cutoff energy. $\Omega$ is given by
 \begin{equation*}
 \begin{split}
 \Omega=\frac{1}{2} \Biggl\{\Delta\omega+i \gamma_\mathrm{X} -\frac{\vec{P}^2}{4 M_\mathrm{T}}-\frac{p_\mathrm{F}^2}{4m_\mathrm{e}} + s\times\quad \quad \quad \quad \quad   \\  \sqrt{\left[ \Delta\omega+i \gamma_\mathrm{X}-\frac{(p_\mathrm{F}+P)^2}{4m_\mathrm{e}} \right]\left[\Delta\omega+i \gamma_\mathrm{X}-\frac{(p_\mathrm{F}-P)^2}{4m_\mathrm{e}} \right]} \Biggr\},
 \end{split}
 \end{equation*}
 where $s=\mathrm{sign}(\Delta\omega-p_\mathrm{F}^2/4m_\mathrm{e}-q^2/4m_\mathrm{e})$. The many-body scattering amplitude $\Gamma$ does not depend separately on $U$ and $\Lambda$, but only on their combination $\epsilon_\mathrm{T}$. As a result it is convenient to treat $\epsilon_\mathrm{T}$ as independent parameter of the model instead of them. Electron-exciton complex represents a simplified model for a trion and  $\epsilon_\mathrm{T}$ is the corresponding binding energy. Excitonic self-energy is connected with the $\Gamma$-vertex as follows \begin{equation}
 \Sigma(\omega,0)=\sum_{\vec{P}} n_\mathrm{F}(\xi_{\vec{P}}^\mathrm{e}) \Gamma(\omega+\xi_{\vec{P}}^\mathrm{e},\vec{P}).
 \label{Sigma0}
 \end{equation}
 Equations (\ref{Pi0}) and (\ref{Sigma0}) are the counterparts of (\ref{Pi}) and  (\ref{Sigma}). The latter reduce to them in the limit of vanishing magnetic field, as we prove in the next section. 
 
\section{C. Recovering the zero-magnetic-field limit}\label{Appendix C}
Here we prove that expressions for the kernel (\ref{Pi}) and self-energy (\ref{Sigma}) reduce to their counterparts (\ref{Pi0}) and (\ref{Sigma0}) in the limit of vanishing magnetic field. In this limit, as well as for high Landau levels $n$, the reconstruction of the electronic spectrum into Landau levels  can be presented as a coarsing of momentum space. All states from $p_n$ to $p_{n+1}$ quench to a single Landau level $n$, where $p_n^2/2 m_\mathrm{e}=\hbar \omega_\mathrm{B} (n+1/2)$. The number of states inside the corresponding ring is equal to the Landau level degeneracy $N_\mathrm{B}$. The set of wave functions can be approximated by plane waves  $|n,l\rangle\rightarrow e^{i \vec{p}_\mathrm{e} \vec{r}}$ with fixed absolute value of momentum $p_\mathrm{e}=p_n$. The summation over Landau orbits $l$ corresponds to the integration over states in the ring, while all rings span the full momentum space 
 \begin{equation}
 \label{Sum}
 \sum_{l}\left(\cdot \cdot \cdot \right)=N_\mathrm{B}\langle \cdot \cdot \cdot \rangle_{\vec{p}_\mathrm{e}},  \quad \sum_{nl}\left(\cdot \cdot \cdot \right) \rightarrow \sum_{\vec{p}_\mathrm{e}}\left(\cdot\cdot\cdot\right). 
 \end{equation}
Here $\langle \cdot\cdot\cdot\rangle_{\vec{p}_\mathrm{e}}$ denotes averaging over direction of momenta $\vec{p}_{\mathrm{e}}$. 

To prove the reduction $\Pi_n(\omega)\rightarrow \Pi(\omega,P_n)$ resulting in $\Gamma_n(\omega)\rightarrow \Gamma (\omega,P_n)$ we at first rewrite (\ref{Pi0}) as follows
\begin{equation}\label{Pi02}
\Pi(\omega,\vec{P})= 2\pi \sum_{\vec{p}_\mathrm{e}\vec{p}_\mathrm{X}}\frac{(1-n_\mathrm{F}(\epsilon_{\vec{p}_\mathrm{e}}))  \Phi(P, p_\mathrm{e},p_\mathrm{X})}{\Delta \omega^+-\frac{\vec{p}_\mathrm{e}^2}{2m_e}- \frac{\vec{p}_\mathrm{X}^2}{2 m_\mathrm{X}}+\epsilon_\mathrm{F}}.
\end{equation}
Here we have introduced the form-factor as follows
\begin{equation} 
\label{Phi0}
\Phi(P, p_\mathrm{e},p_\mathrm{X})=2\pi \langle \delta(\vec{P}-\vec{p}_\mathrm{e}-\vec{p}_\mathrm{X})\rangle_{\vec{p}_\mathrm{e}, \vec{p}_\mathrm{X}}
\end{equation} In the limit of vanishing magnetic field reduction of (\ref{Pi}) to (\ref{Pi02}) requires 
 \begin{equation}\label{PhiConnection}
 l_\mathrm{B}^2 \Phi^{n'}_{n}\left(\frac{p^2_\mathrm{X} l_\mathrm{B}^2}{2}\right)\rightarrow\bar{\Phi}(P_{n'},p_{\mathrm{e},n}, p_\mathrm{X}).
 \end{equation} 
 To prove this relation we recall the definition of the form-factor $\Phi^{n'}_n(x)$ 
\begin{widetext}
	\begin{equation}
	\delta_{n'' n'}^{l'' l'} \Phi^{n'}_n\left(\frac{p_\mathrm{X}^2 l_\mathrm{B}^2}{2}\right)=\sum_{l} \langle F^{l'' l}_{n'' n}(-\vec{p}_\mathrm{X}) F^{l l'}_{n n' }(\vec{p}_{\mathrm{X}})\rangle_{\vec{p}_\mathrm{X}}\rightarrow  (2\pi \hbar)^2 \delta(\vec{P}_{n''}-\vec{P}_{n'} )  \times (2\pi \hbar)^2  N_\mathrm{B} \langle  \delta(\vec{P}_{n'}-\vec{p}_{\mathrm{e},n}-\vec{p}_\mathrm{X})   \rangle_{\vec{p}_\mathrm{X},\vec{p}_\mathrm{e}}. 
	\end{equation} 
\end{widetext}
The first multiplier in the last term corresponds to $\delta_{n'' n'}^{l'' l'}\rightarrow (2\pi \hbar)^2 \delta(\vec{P}_{n''}-\vec{P}_{n'} )$. The second multiplier ensures the connection (\ref{PhiConnection}).  

The connection between the self-energies (\ref{Sigma}) and (\ref{Sigma0}) in the limit of vanishing magnetic field can be proven as follows.
\begin{widetext}
\begin{equation}
\Sigma(\omega,0)=N_\mathrm{B}\sum_{n} \nu^\mathrm{e}_n \bar{\Gamma}_{n} (\omega+\epsilon^\mathrm{e}_n) \rightarrow N_\mathrm{B}\sum_{n} n_\mathrm{F}(\epsilon_{P_n}^\mathrm{e}) \Gamma (\omega+\epsilon^\mathrm{e}_{P_n}, P_n) \rightarrow \sum_{\vec{P}} n_\mathrm{F}(\epsilon_{\vec{P}}^\mathrm{e}) \Gamma(\omega+\epsilon_{\vec{P}}^\mathrm{e},\vec{P}).
\end{equation} 
\end{widetext} 

\section{D. Strong magnetic field limit}
In the limit of the strong magnetic field $\epsilon_\mathrm{T}\ll\hbar \omega_\mathrm{B}$ only few Landau levels are important and need to be taken into account. Their number can be estimated as $N_\mathrm{L}\approx\epsilon_\mathrm{X}/\hbar \omega_\mathrm{B}$, where 
the binding energy of excitons $\epsilon_\mathrm{X}$ is the energy scale at which excitons cannot be considered as structureless quasiparticles anymore. Taking into account that in experiments exciton and trion binding energies differ by an order of magnitude  $\epsilon_\mathrm{X}/\epsilon_\mathrm{T}\sim 10$ we can estimate the number of
 relevant Landau levels as $N_\mathrm{L}\approx\epsilon_\mathrm{X}/\hbar \omega_\mathrm{B}\sim 5$. 
 In that case the self-energy $\Sigma(\omega,\vec{p=0})$ of excitons at zero momentum  can be rewritten as

\begin{equation}
\label{ExcitonXFew}
\Sigma=-\frac{\hbar \omega_\mathrm{B}}{2} \sum_n  \frac{\alpha_\mathrm{X} \nu^\mathrm{e}_n  }{1+\alpha_\mathrm{X} \sum_{n'}\bar{\nu}^\mathrm{e}_{n'} I^{n'}_{n}[2(\frac{\Delta \omega}{\hbar \omega_\mathrm{B}}+n-n')]},
\end{equation}
where $\alpha_\mathrm{X}=N_\mathrm{X}|U|$ is the dimensionless coupling constant with $N_\mathrm{X}=m_\mathrm{X}/2\pi \hbar^2$ to be the excitonic density of states. It appears instead of the reduced density of states of electron and exciton $N_\mathrm{T}=\mu_\mathrm{T}/2\pi \hbar^2$ since kinetic energy of electrons is quenched.  The function $I_{n n'}[z]$ is defined as follows 
\begin{equation}
I_{n n'}[z]=\int_0^{\infty}dx  \frac{\Phi^{n'}_n(x)}{z-x+ i\delta}.
\end{equation}
Its imaginary part is given by $I''_{n n'}[z]=-\pi\Theta(z) \Phi^{n'}_n(z)$, while the real part $I'_{n n'}[z]$ for the first three Landau levels is presented in the following Table.

Here $\text{Ei}(z)$ is the exponential integral function. The expression of $I'_{n n'}[z]$ for higher Landau level indices $n$ and $n'$ has the same functional form.

\begin{figure}[b]
	\label{Fig5}
	\vspace{-2 pt}
	\includegraphics[width=0.97\columnwidth]{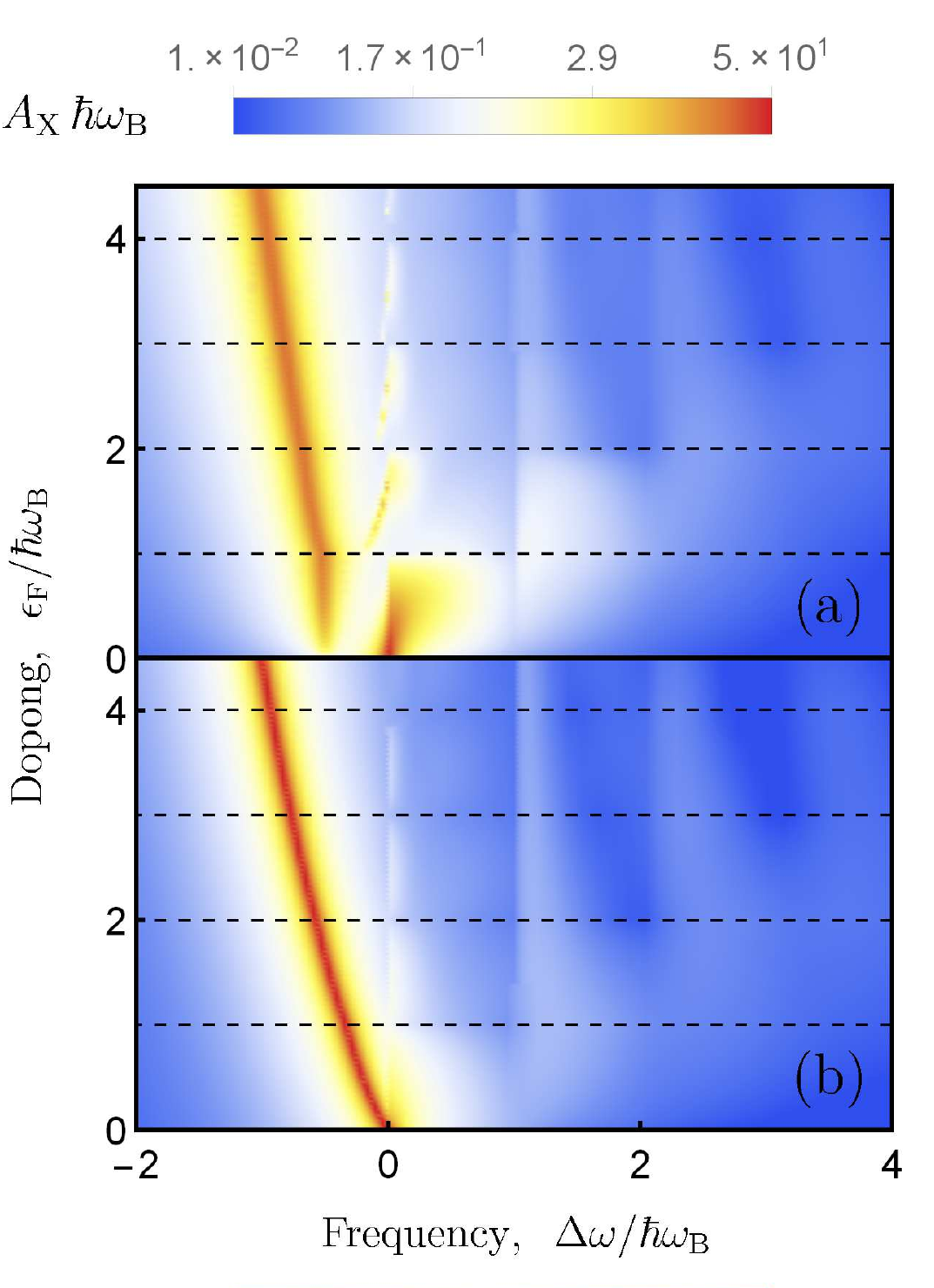}
	\vspace{-2 pt} 
	\caption{Analytical results for the spectral function of excitons  $A_\mathrm{X}(\omega,\vec{p}=0)$ at $\alpha_\mathrm{X}=0.4$ and $N_\mathrm{L}=5$. Full ladder approximation $(\hbox{a})$ and the second-order perturbation theory $(\hbox{b})$. The perturbation theory does not capture the splitting of redshifted attractive exciton-polaron branch, but well describes the appearance and doping dependence of exciton-cyclotron resonances.}
\end{figure}

In this spectral function of excitons at zero momentum  $A_\mathrm{X}(\omega,0)$ depends on dimensionless frequency $\omega/\hbar \omega_\mathrm{B}$ and doping $\nu_\mathrm{e}=\epsilon_\mathrm{F}/\hbar \omega_\mathrm{B}$, as well as on dimensionless coupling constant $\alpha_\mathrm{X}$ and number of involved Landau levels $N_\mathrm{L}$. For $\alpha_\mathrm{X}=0.4$ and $N_\mathrm{L}=5$ the spectral function $A_\mathrm{X}(\omega,0)$ is presented in Fig.~5~(a). Compared to results described in the main text, the few Landau level approximation very well captures both splitting of the exciton level into attractive and repulsive exciton-polaron branches. It also captures the appearance of features corresponding to exciton-cyclotron resonances. It correctly predicts frequencies of all resonance and doping dependencies of their amplitudes.     

To investigate the capability of the perturbation theory we expand the self-energy (\ref{ExcitonXFew}) in the coupling constant $\alpha_\mathrm{X}$ up to the second order. The first order term $\Sigma^{(1)}(\omega)=-\alpha_\mathrm{X} \nu_\mathrm{e} \hbar \omega_\mathrm{B}/2$ represents the shift of the energy of excitons due to their interactions with electrons. The second order term is given by
\begin{equation*}
\Sigma^{(2)}(\omega)=\alpha_\mathrm{X}^2 \frac{\hbar \omega_\mathrm{B}}{2}\sum_{n n'} \nu^\mathrm{e}_n \bar{\nu}^\mathrm{e}_{n'} I_{n n'}\left[2\left(\frac{\Delta \omega}{\hbar \omega_\mathrm{B}}+n-n'\right)\right],
\end{equation*}
and represents the effect of electronic transitions between Landau levels. The corresponding frequency and doping dependence of the spectral function of excitons $A_\mathrm{X}(\omega,0)$ is presented in Fig.~5~(b). The perturbation theory very well captures the appearance and doping dependence of the exciton-cyclotron resonances. Nevertheless, it is not sufficient to describe the splitting of redshifted attractive exciton-polaron branch.    

\begin{widetext}\begin{center}
		\begin{tabular}{|c|c|c|c|}
			\hline
			$\quad I'_{n n'}[z]\quad$ &\quad $0$\quad &\quad $1$ \quad&\quad $2$\quad \\[3pt]\hline
			$0$ & $e^{-z} \text{Ei}(z)$, & $e^{-z} z \text{Ei}(z)-1$, & $\frac{e^{-z} z^2 \text{Ei}(z)-z-1}{2} $  \\[3pt]\hline$1$ &	$e^{-z} z \text{Ei}(z)-1$ &	$e^{-z} (z-1)^2 \text{Ei}(z)-z+1$ & $\frac{e^{-z} (z-2)^2 z \text{Ei}(z)-z^2 +z +2 }{2}$  \\[3pt]\hline $2$&\quad
			$\frac{e^{-z} z^2 \text{Ei}(z)-z-1}{2} $ &\quad 	$\frac{e^{-z} (z-2)^2 z \text{Ei}(z)-z^2 +z +2 }{2}$ \quad&\quad$\frac{(z^2-4z+2)^2 e^{-z}\text{Ei}(z)- z^3+7z^2-14z +6 }{4}$ \\\hline
		\end{tabular}
	\end{center}
\end{widetext}

\end{document}